\documentstyle[12pt]{article}
\newcommand{\bce}{\begin{center}}
\newcommand{\ece}{\end{center}}
\newcommand{\bea}{\begin{eqnarray}}
\newcommand{\eea}{\end{eqnarray}}
\newcommand{\be}{\begin{equation}}
\newcommand{\ee}{\end{equation}}
\newcommand{\bd}{\begin{displaymath}}
\newcommand{\ed}{\end{displaymath}}
\newcommand{\bit}{\begin{itemize}}
\newcommand{\eit}{\end{itemize}}
\newcommand {\ben}{\begin{enumerate}}
\newcommand{\een}{\end{enumerate}}
\newcommand{\bfx}[1]{\mbox{\boldmath $#1$}}

\begin{document}
\bce
{\bf Nuclear Polarization Corrections for the S--Levels of
Electronic \newline
\vspace{-0.8cm}
\center
and
Muonic Deuterium}

\vspace{0.5cm}
Yang Lu and R.~Rosenfelder

\vspace{0.3cm}
Paul Scherrer Institute, CH-5232 Villigen PSI, Switzerland.
\ece

\vspace{1cm}
\begin{abstract}
\noindent
We calculate the second-order corrections to the
atomic energy level shifts in ordinary
and muonic deuterium due to virtual excitations of the deuteron
which are
important for ongoing and planned precise experiments in these systems.
For light atoms a method can be used in which
the shift is expressed as integrals over the longitudinal
and transverse inelastic structure functions of the nucleus.
We employ the structure functions arising from separable NN
potentials of the Yamaguchi and Tabakin form which can be evaluated
analytically. Special emphasis is put on gauge
invariance which requires a consistent
inclusion of interaction currents and seagull terms.
The effect of the D-wave component of the deuteron is investigated
for the
leading longitudinal contribution. We also estimate the shift for
pionic
deuterium.
\end{abstract}
\newpage

\noindent
{\bf 1.} Recently the isotope shift of the $2S-1S$ transition in
electronic
hydrogen and deuterium has been measured
with a thirty-fold increase in accuracy compared to previous
experiments
\cite{SLWH} and prospects are good to increase the present accuracy
of $37$ ppb (or $22$ kHz) by another order of magnitude. At this
level
not only various QED corrections and the finite size of the deuteron
are important but also virtual excitations of the deuteron (the
so-called
nuclear polarization) cannot be neglected anymore. Using a rough
square-well model of the deuteron and the closure approximation
the 1S shift due to virtual Coulomb excitation
 was estimated to be about $- 19$ kHz \cite{PLH}
which is just the present experimental uncertainty. In light and heavy
muonic atoms the nuclear polarization shift generally limits the
accuracy
with which nuclear sizes can be extracted since a reliable calculation
of these
corrections requires knowledge of the whole nuclear spectrum \cite{Ro1}.
The deuteron is unique among all nuclei in that
this information is available: the
quantum-mechanical 2-body problem is solvable and realistic NN
potentials
describe bound and scattering states rather well. Therefore, unlike
the usual
case, nuclear polarization corrections are {\it calculable}
for the deuteron -- presumably with an accuracy at the percent level.

It is the purpose of the present note to evaluate this shift within a
more
realistic model for the deuteron, avoiding the use of the closure
approximation and including also transverse excitations
which have been neglected up to now. For the purpose of a planned
experiment
at PSI with muonic hydrogen \cite{Zav} which may be extended to
deuterium
\cite{Sim}
we also evaluate the corresponding shifts in
the muonic case. It should be noted that this
does {\it not} amount to a simple
rescaling of the electronic results: since nuclear excitations are
of the order
of  a few MeV the electron is higly relativistic in the
virtual state whereas
the muon can be treated nonrelativistically to a good approximation.
 This leads
to different weighting of the individual excitations and
therefore to quite
different results.

\vspace{0.5 cm}
\noindent
{\bf 2.} We will calculate the nuclear polarization shifts
in the atomic S-states of the lepton. These are more difficult
to evaluate
than the corresponding shifts in higher orbits where only the
longest range
multipole is of importance and therefore the only nuclear
structure information needed is the electric dipole
polarizability of the nucleus. In contrast, many multipoles
contribute to
virtual excitations from atomic S-states since there is an
overlap between
leptonic and nuclear wavefunctions. In light nuclei,
however, the relevant scales
(Bohr radius vs. nuclear radius) are vastly different so that
the S-wave lepton wavefunction can be considered as
approximately constant
over the nuclear volume $ \phi_{n0}({\bf x}) \simeq \phi_{\rm av} $ :
the lepton then just acts as a static source with four-momentum
$k = (m,{\bf 0})$ .

We will evaluate the S-wave nuclear polarization  shifts
along the lines of ref. \cite{Ro2} , i.e. not by a multipole
decomposition but by integrating over the inelastic structure
functions of the nucleus.
The diagrams which contribute to nuclear
polarization are shown in Fig. 1.
Note that we need the ``seagull'' contribution of Fig. 1 (c)
for gauge
invariance in a nonrelativistic sytem like the nucleus.
Under the mentioned simplifications one obtains
\be
\Delta \epsilon_{n0} = - \> \frac{(4 \pi \alpha )^2}{m}
|\phi_{\rm av}|^2
 \> {\rm Im} \> \int
\frac{d^4 q}
{(2 \pi)^4} \> \> t_{\mu \nu} (q,k)
\> D^{\mu \rho} (q) \> D^{\nu \tau} (-q) \> \bar T_{\rho \tau} (q,-q)
\ee
where
\be
t_{\mu \nu} (q,k) = \frac{k \cdot q \> g_{\mu \nu} +
(k-q)_{\mu}k_{\nu}
+ k_{\mu}(k-q)_{\nu} }{ (k-q)^2 - m^2 + i \epsilon}
\ee
is the leptonic tensor, $\alpha = e^2 = 1/137.036$ the
fine-structure constant
and $m$ is the lepton mass. Furthermore,
$D^{\mu \rho} (q)$ denotes the photon propagator and
$ \bar T_{\rho \tau} (q,-q)$
the {\it forward} virtual nuclear Compton amplitude.
To be more precise,
the latter is that part of the full Compton amplitude in
which the nucleus
is in an excited intermediate state. It can be
expressed in terms of its imaginary part, i.e. by the inelastic
longitudinal and transverse structure functions
\be
S_{L/T} = \sum_{N \ne 0} \> \delta(\omega' + E_0 - E_N) \> |<\psi_N|
{\cal O}_{L/T} |\psi_0>|^2 \> .
\ee
Here $\omega' = \omega - q^2/4M$ is the internal excitation energy,
$E_0 < 0 $ the ground state energy and
${\cal O}_{L/T}$ are the operators for longitudinal and transverse
excitations
repectively.
In Coulomb gauge one obtains \cite{Ro2}
\bea
\Delta \epsilon_{n0} = - \>  8 \alpha^2 \> \bar R_{n0} \>
|\phi_{n0}(0)|^2\> \int_0^{\infty}
dq \>   &\biggl \{& \int_0^{\infty} d\omega \> \biggl [
 \> K_L (q,\omega) \> S_L(q,\omega) \> +
\> K_T (q,\omega) \> S_T(q,\omega) \> \biggr ] \nonumber \\
\> &+& \> R_S(q) \biggr \} \> f^2 (q) \> .
\eea
Here $ \bar R_{n0}$ is a correction
factor for the variation of the leptonic
wave function over the nucleus, $q$ is the magnitude of the
three-momentum
transfer and $\omega$ the energy transfer to the nucleus.
 The kernels $K_{L,T} (q,\omega)$  are given
in the appendix of ref. \cite{Ro2} for fully relativistic
kinematics of the lepton.
 $R_S(q)$ is the contribution from
the {\it internal} seagull \footnote{The center-of-mass
seagull term is part
of the two-photon recoil correction \cite{GrYe}.
Note also that the part of the
Compton amplitude in which the nucleus remains in its ground
state has already been included by solving
the bound state problem for the lepton with a static potential.}.
Finally, $f(q) = (1 + q^2/(0.71 {\rm GeV}^2))^{-2}$
describes the electromagnetic formfactor of the nucleon.
Actually eq. (4) not only holds in the Coulomb gauge
but is gauge invariant
provided $q^{\mu} \bar T_{\mu \nu}(q,-q) = 0$. This in turn
requires current
conservation and special conditions for the seagull
term \cite{FrRo,Are}
which will be discussed below.

\vspace{0.5cm}

\noindent
{\bf 3.} Unlike ref. \cite{Ro2} where a phenomenological
model for the structure
functions of $^{12}$C had to be used, the deuteron
allows for a consistent
{\it calculation}
of these quantities after a nucleon-nucleon interaction
has been chosen.
For simplicity we take a separable potential of the form
($M$ is the nucleon mass)
\be
V({\bf p},{\bf p'}) = - \frac{\lambda}{M} \> g(p) \> g(p') \>.
\ee
This is not realistic in the modern sense (it lacks the
one-pion exchange tail
and all other complications of the NN force) but it gives a
fairly good
description of the low-energy NN interaction which
should be sufficient
for the present accuracy of isotope shift experiments. Most
important
for the present purposes it
allows for an analytic evaluation of the structure functions.
For example,
the longitudinal structure function is obtained as \cite{Ro3}
\be
S_L(q,\omega) = \int d^3 p \> |\psi_0({\bf p -  q}/2) |^2
\> \delta ( \omega' + E_0 - \frac{p^2}{M} ) \> + \>
\frac{\lambda M}{\pi} \>
{\rm Im} \left ( \frac{ I^2(\omega' + E_0,q)}{1 + \lambda \>
C(\omega' + E_0 )}
\right ) \> .
\ee
Here the functions $C(E)$ and $I(E,q)$ are given by
\bea
C(E) &=& \int d^3p \> \frac{g^2(p)}{M E - p^2 + i \epsilon} \\
I(E,q) &=& \int d^3p \> \psi_0({\bf p - q}/2) \>
\frac{g(p)}{M E - p^2 + i \epsilon} \> \> .
\eea
Note that the internal charge operator is
\be
\hat \rho({\bf q}) =  \exp ( \frac{i}{2} {\bf q} \cdot
\hat{\bf r} )
\ee
where the factor $1/2$ arises from the transformation to
internal coordinates
(a ``hat'' indicates an operator in the nuclear Hilbert space).
The ground state wave function is simply given by
\be
\psi_0({\bf p}) = N \> \frac {g(p)}{p^2 - M E_0}\> .
\ee
Also note that
the first term in eq. (6) is the impulse approximation
to the structure function whereas the last one describes the
final-state
interaction. Due to the simple form of eq. (5) it only
acts in states with
angular momentum zero which should be a good approximation for
low-energy
processes.
For the Yamaguchi choice \cite{Yam}
\be
        g_Y(p) = \frac{1}{p^2 + \beta^2}
\ee
all integrals can be performed analytically.
Details will be given elsewhere.

The transverse nuclear polarization shift receives different
contributions :
first, we have the contribution from the
transverse (with respect to ${\bf q}$ )
part of the convection current
\be
\hat {\bf J}^{({\rm conv} )} ({\bf q}) = \frac{1}{2M} \>
\left [ \> \hat {\bf p} \>,
 \> \exp( \frac{i}{2} {\bf q} \cdot \hat {\bf r} ) \> \right ]
\ee
which does not have a final-state interaction
term since all excited states
necessarily have at least angular momentum one. Second,
the spin current
\be
\hat {\bf J}^{({\rm spin})} ({\bf q}) = \frac {i}{2 M}
\left [ \> \mu_p \>
(\hat {\bfx \sigma}_p \times {\bf q} ) \exp(
\frac{i}{2} {\bf q} \cdot
\hat{\bf r} )
+ \mu_n ( \hat {\bfx \sigma}_n \times {\bf q} )
\exp( -  \frac{i}{2} {\bf q} \cdot \hat{\bf r} ) \> \right ]
\ee
involves the magnetic moments and spins of neutron and proton and
gives rise to spin-flip excitations.  Since
the procedure of calculating the transverse structure
with these currents
is similar to the longitudinal case
the explicit expressions will not be given here. Note that
the operators
(9), (12) and (13) obey current conservation in the form
\be
<{\bf p'}| \> \left [ \hat T, \hat \rho({\bf q}) \right ] -
{\bf q} \cdot \left (
\hat {\bf J}^{({\rm conv})} ({\bf q})\> + \hat {\bf J}^{({\rm spin})}
({\bf q})
\> \right ) | {\bf p}>  = 0
\ee
where $\hat T$ is the kinetic energy operator.

\vspace{0.5 cm}
\noindent
{\bf 4.} A separable potential of the type (5)
is nonlocal and equivalent to a momentum-dependent potential
\be
\hat V = - \frac{\lambda}{M} \> \int d^3 r \> d^3 x \>
g \left ( {\bf r} + \frac{1}{2}{\bf x} \right) \>
g \left ( {\bf r} - \frac{1}{2}{\bf x} \right) \>
\> \exp(-i \hat {\bf p} \cdot {\bf x}/2)
\> |{\bf r}> <{\bf r}| \>
\exp(-i \hat {\bf p} \cdot {\bf x}/2) \> .
\ee
Minimal coupling $ \hat {\bf p}_p \longrightarrow \hat {\bf p}_p -
e {\bf A}( {\bf r}_p) $
in the two-body Hamiltonian $ \hat H = \hat T + \hat V$
then produces a power series in the electromagnetic coupling
constant $e$
\be
\hat H = \hat H \bigl |_{A=0}  \> - \> e \> \int d^3y \hat
J_{\mu} ({\bf y})\>
A^{\mu}({\bf y})
\> + \> \frac{e^2}{2} \int d^3y \> d^3z \> A_i({\bf y}) \> A_j({\bf z})
\> \hat B_{ij}({\bf y},{\bf z})\> + \> ...
\ee
The linear terms due to the potential generate an interaction current
\be
 e \> \Delta \hat J_k ({\bf y}) = - \frac{\delta \hat V}{\delta
A_k({\bf y})}
\biggl |_{A=0}
\ee
whereas the second-order terms give rise to the interaction seagull
\be
e^2 \> \Delta \hat B_{ij}({\bf y},{\bf z}) = \>
\frac{ \delta^2 \hat V}
{\delta A_i({\bf y}) \>\delta A_j({\bf z}) } \biggl |_{A=0} \> .
\ee
With the separable potential (15) one obtains the following expression
for the matrix element of the interaction current
\be
<{\bf p'}| \Delta \hat {\bf J}({\bf q}) | {\bf p}> = -
\frac{\lambda}{2 M}
\> \left ( \nabla_p + \nabla_{p'} \right ) \> \int_0^1 du \>
g \left ({\bf p} - u \frac{\bf q}{2} \right ) \>
g \left({\bf p'} + (1-u) \frac{\bf q}{2} \right)
\ee
which satisfies
\be
<{\bf p'}| \> \left [ \> \hat V, \hat \rho({\bf q})\> \right ] -
{\bf q} \cdot
\Delta \hat J({\bf q})\> | {\bf p}>  = 0
\ee
so that all together our currents are conserved under
the time-evolution
of the full Hamiltonian.
\par\noindent
Similarly one obtains
the following matrix element of the interaction seagull
\bea
&<&{\bf p'}| \> \Delta \hat B_{i j}({\bf q})
\> | {\bf p}> \equiv <{\bf P}_f = {\bf 0},{\bf p'} | \>
\int d^3 x \exp(i {\bf q} \cdot {\bf x}) \>  \Delta
\hat B_{ij} ({\bf x},
{\bf 0})
\> | {\bf p}, {\bf P}_i = {\bf 0} \> > \nonumber \\
&=& - \frac{\lambda}{4 M }
\> \left ( \nabla_p + \nabla_{p'} \right )_i  \>
\left ( \nabla_p + \nabla_{p'} \right )_j
\> \int_{-1}^{+1} du \> ( 1- |u|) \>
g \left({\bf p} - u \frac{\bf q}{2} \right) \>
g \left({\bf p'} - u \frac{\bf q}{2} \right)
\eea
which comes in addition to the kinematical (internal) seagull
$ \delta_{ij}/2 M$.

It is well known \cite{FrRo,Are} that
for full gauge invariance of the Compton amplitude
a relation between
the interaction current and the interaction seagull
is needed. In the present case where no overall momentum is
transferred
to the target it reads
\be
<{\bf p'}| \>  \> [ \> \hat \rho^{\dagger}({\bf q}),
\Delta \hat J_k ({\bf q}) \> ]
\> \> | {\bf p} > = q \> <{\bf p'}| \> \Delta \hat B_{3 k}({\bf q})
\> | {\bf p}> \> .
\ee
With the interaction current and interaction seagull as given
in eqs. (18) and (21) this relation can be shown to be fulfilled.
Thus
the nuclear polarization shift is gauge invariant
if the interaction terms in current and seagull are taken
into account
consistently. The seagull contribution in eq. (4) is now
explicitly given by
\be
R_S(q) = \frac{1 + \kappa(q)}{8 m M} \> \left ( \frac{1}{q} -
\frac{1}{\sqrt{m^2 + q^2}} \right )
\ee
where
\bea
\kappa(q) &=& \frac{4 \lambda}{q^2} \> \int_0^1 du \>
\frac{1}{u^2}
\left [ h^2 ( 0 ) \> - \> h^2( u q ) \right ] \\
h(x) &=& \int d^3p \> \> \psi_0(p) \> g \left (
{\bf p} - \frac{{\bf x}}{2} \right )
\> .
\eea

\vspace{0.5 cm}

\noindent
{\bf 5.} We now turn to the numerical results of our calculation.
We have evaluated the nuclear polarization shift consistently with the
Yamaguchi separable form (5) using the value $ \beta = 286 $ MeV.
We have checked numerically that our structure function fulfills the
non-energy-weighted sum rule
\be
\int_0^{\infty} d\omega \> S_L(q,\omega) \> = \> 1 \> - \> F_0^2(q)
\ee
to better than 1 part in $10^5$.
Here $F_0(q)$ is the elastic formfactor calculated directly from the
ground state wave function (10).
The electric dipole polarizability (which is {\it not} the only relevant
quantity for S-wave shifts)
was found to be $0.613 $ fm$^3$ compared to
experimental values of $ 0.61 - 0.70 $ fm$^3$ \cite{FrFa}. The point
rms-radius in this model is $1.92$ fm compared to the experimental
value
$1.96 $ fm. This shows that the simple Yamaguchi parametrization
describes the deuteron properties and therefore the low-energy
 triplet NN-interaction reasonably well.
For the spin-flip excitations one also needs the parameters in the
singlet channel. Again following Yamaguchi we assume
$\beta_s = \beta_t$ and determine the corresponding strength
parameter from the singlet scattering length $a_s = - 23.69$ fm.

For non-relativistic point hydrogen wave functions one has
\be
|\phi_{n0}(0)|^2 = \frac{1}{\pi a_B^3} \> \frac{1}{n^3}
\ee
where $ a_B = 1/(\alpha m_{\rm red})$ is the Bohr radius and
$m_{\rm red}$
the reduced mass of the lepton. Writing
\be
\Delta \epsilon_{n0} \> = \> \frac{\bar R_{n0}}{n^3} \Delta \bar
\epsilon \> .
\ee
the shift $\Delta \bar \epsilon$ is then independent of the atomic
state.
We have evaluated the double integral
in eq. (4) by Gauss-Legendre numerical
integration with up to $3 \times 72$ points.
Our results for the different contributions and for the total
$\Delta \bar \epsilon$ are listed in Table 1. It should be emphasized
that the integrand
from the transverse convection current has a $ 1/q$ - divergence for
small $q$ which , however, is exactly cancelled by the seagull
due to the
gauge condition
for the two-photon operator. This can also be seen from
eqs. (4) and (23) : at low $q$ the kernel $K_T(q,\omega)$
behaves like
to $ - 1/ 4 m q \omega \> \> $ \cite{Ro2} and
Siegert's theorem tells us that
\be
\int_0^{\infty} d\omega \> \frac{1}{\omega} \> S_T(0,\omega) =
\frac{1 + \kappa}{2 M}
\ee
where $\kappa \equiv \kappa(0) $ is the dipole enhancement factor
($ \kappa = 0.176$ for the Yamaguchi potential). However, the
resulting contribution to the energy shift is
exactly opposite in sign to the
$q \to 0$-limit of the seagull contribution (23). We have checked
numerically
that our transverse structure function fulfills the sum rule (29) to
sufficient
accuracy. Note that in ref. \cite{Ro2} the seagull was represented by
eq. (29) for {\it all} $q$; we have found numerically
that the extra $q$-dependence only gives a tiny
contribution to the full shift.

In Table 1 we therefore give only the combined
contribution of transverse convection and seagull excitations.
It is seen that
it is bigger in electronic deuterium than in muonic deuterium because
the electron velocity is higher in the first case.
As the spin
current contribution vanishes for $q = 0$ (see eq. (13)) it
can be given separately. However, numerically it turns out to be of
no great
importance. The same can be said of the interaction terms
which nearly
cancel the spin contribution. The smallness of the interaction terms
 is welcome since
the nonlocality of the Yamaguchi separable
potential is somehow artificial and only partly simulates exchange
current effects.
It should be kept in mind
that the individual contributions are
gauge-dependent and that only the
total $\Delta \bar \epsilon$ is a meaningful physical quantity.
The size
of the transverse and seagull terms , however,
indicates
the errors one usually makes if only the longitudinal
excitations are taken into account. As to the numerical accuracy,
we have
checked that the results in Table 1
are accurate to one part in the last digit.

In order to estimate the model dependence of these results we
also have
calculated the dominating longitudinal contribution for the
Tabakin separable potential \cite{Tab} which
describes both attraction at low energies
and repulsion at higher energies in
the S-wave phase shift. The principal value integrals in
eqs. (7) and (8)
have now been performed numerically which considerably increased
the computing
time. Again the sum rule was checked and
an electric dipole polarizability of $0.623$ fm$^3$ was obtained.
The values for the shift given in Table 2 are estimated to have
an accuracy
of about three parts in the last digit.
Despite
the different functional parametrization of $g(p)$ the
result for $\Delta \bar \epsilon$ is very close to the Yamaguchi
one which
shows that only
low-energy properties of the NN-interaction are important for the
nuclear
polarization shift.

Finally we also have investigated the influence of the
D-state admixture
in the deuteron by using the Yamaguchi separable potential with
tensor force
\cite{YaYa}
\be
g({\bf p}) = g_Y(p) + \frac{1}{\sqrt{8}} \> T_Y(p) \> S_{pn}({\bf p})
\ee
with
\bea
T_Y(p) &=& - \frac{t p^2}{(p^2 + \gamma^2)^2} \\
S_{pn}({\bf p}) &=&  3 \> \frac{{\bfx \sigma}_p \cdot {\bf p} \>
{\bfx \sigma}_n
\cdot {\bf p}}{p^2} \> - \> {\bfx \sigma}_p \cdot {\bfx \sigma}_n \> .
\eea
The original parameter values of ref. \cite{YaYa} lead to an asymptotic
D/S ratio
of $0.0285$ which is quite reasonable when compared with modern values
\cite{ErRo}. The dipole polarizability is calculated to be $0.625$
fm$^3$.
Since the algebra including the tensor force is more involved
the sum rule check (to one part in $10^3$ ) is nontrivial.

To convert these numbers to the actual nuclear polarization shifts
for the
S-levels we need the finite size correction factors $\bar R_{n0}$ .
Using the approximate atomic wavefunctions of ref. \cite{Fri}
one obtains in first order in the ratio nuclear radius/Bohr radius
\be
\bar R_{n0} \simeq 1 - 3.06 \frac {<r^2>^{1/2}}{a_B} \> .
\ee
The numerical factor in this equation was determined by evaluating
the ratio of various moments of the charge distribution
with the Yamaguchi ground state wavefunction.
Eq. (33) gives {\it n-independent} correction factors
$\bar R^{(\mu)} = 0.9793 $ and $\bar R^{(e)} = 0.99989 $.
Of course, on the
present level of accuracy one can practically
neglect these correction factors.

We estimate the accuracy of our theoretical predictions
in the following
(rather conservative) way:
the accuracy of the calculated longitudinal shift is taken
as three times the
model-dependence shown in Table 2 and we assign a
$20$ \% error to the transverse current contribution
and $50$ \% one to the interaction pieces. Adding these
errors linearly
we therefore arrive at the final result
for the nuclear polarization shifts in electronic and
muonic deuterium
\bea
\Delta \epsilon_{n0}^{(e)} &=& \left ( - 20.5 \pm 1.3 \right ) \>
\frac{1}{n^3} \> \> \> {\rm kHz} \\
\Delta \epsilon_{n0}^{(\mu)} &=& \left ( - 11.6 \pm 0.5 \right ) \>
\frac{1}{n^3}
\> \> \> {\rm meV} \> .
\eea
We also have estimated the nuclear
polarization shift in pionic deuterium
by replacing the muon mass by the pion mass. The longitudinal and the
transverse convection current contribution of the present formalism
should give a reasonable value
also for a heavy spin zero particle because to a good approximation
it can be treated
nonrelativistically with no difference between a
Dirac and a Klein - Gordon description. In this way we obtain
\be
\Delta \epsilon_{n0}^{(\pi)} \simeq - 28  \> \> \frac{1}{n^3}
\> \> \> {\rm meV} \> .
\ee
For $ n = 1 $ this is a factor of two smaller than
the precision aimed at
in an ongoing
experiment at PSI to measure the strong interaction shifts
in pionic hydrogen \cite{Gou}.

If the future isotope shift experiments in
electronic deuterium actually
reduce the experimental accuracy to
about $1 $ kHz \cite{SLWH} it would be
worthwhile to repeat the present calculation with a modern
NN-potential like
the Paris potential \cite{Paris}. Also it may be useful
to employ a general
gauge like the ``$\alpha$-Lorentz gauge'' \cite{Bax} which
interpolates between
Coulomb and Lorentz gauge
thereby demonstrating explicitly the independence
of the numerical results from the gauge parameter $\alpha$.
Finally, before discrepancies between theory and
experiment in the
isotope shifts are taken serious
one should include second-order effects also in the
analysis of
electron-deuteron scattering experiments which extract
the root-mean square
radius of the deuteron. In the case of $^{12}$C similar
discrepancies
between electron scattering data and muonic energy
shifts seem
to disappear \cite{Off} when second-order effects are
taken into account in
the analysis of {\it both} experiments.

\vskip 3 true cm

{\bf Acknowledgements}\\
\vskip 0.3 true cm
We thank Andreas Schreiber for helpful discussions and a critical
reading of the manuscript. We are grateful to Leo Simons and
Pieter Goudsmit for providing us with experimental information.

\newpage

\noindent
{\large \bf Table 1 :}\\
Contributions to the nuclear polarization shift
$\> \> \Delta \bar \epsilon \> \> $
( see eq. (28) ) for electronic (e) and muonic ($\mu$)
deuterium in Coulomb
gauge. The Yamaguchi
S-wave separable potential has been used throughout.
The different
contributions are labeled by L : longitudinal;
T$^{({\rm conv})}$ + S  :
tranverse convection current + seagull; T$^{({\rm spin})}$  :
transverse
spin current; $\Delta$(T + S) : interaction
transverse current + interaction seagull.

\vspace{0.5cm}
\begin{tabular}{|l|r|r|} \hline
 Contribution           & e \ [kHz] \    & $\mu$ \ [meV]\   \\ \hline
 L                      & --18.31        & --11.77  \\
 T$^{({\rm conv})}$ + S & -- 2.25        & -- 0.06  \\
 T$^{({\rm spin})}$     & + 0.33         & + 0.03    \\
 $\Delta$(T + S)        & -- 0.31        & -- 0.02    \\
 total                  & --20.54        & --11.82  \\ \hline
\end{tabular}

\vspace{2 cm}

\noindent
{\large \bf Table 2 :}\\
Longitudinal nuclear polarization shift $\Delta \bar \epsilon$
for different
separable NN-potentials.

\vspace{0.5cm}
\begin{tabular}{|l|r|r|} \hline
 NN-Potential                   & e \ [kHz] \  & $\mu$ \  [meV]\  \\
\hline
 Yamaguchi S-wave \cite{Yam}    & --18.31      & --11.77  \\
 Tabakin S-wave   \cite{Tab}    & --18.54      & --11.92  \\
 Yamaguchi S+D-wave \cite{YaYa} & --18.45      & --11.86  \\  \hline
\end{tabular}

\vspace{2cm}
\noindent
{\large \bf Figure caption}

\vspace{1cm}
\noindent
{\large \bf Fig. 1 :}\\
\\
Second-order contributions to the nuclear polarization shift:
(a) box graph, (b) crossed graph, (c) seagull graph.

\end{document}